\documentstyle[12pt]{article}

\makeatletter
\def\@cite#1#2{$^{\hbox{\scriptsize{#1\if@tempswa , #2\fi}}}$}
\makeatother

\newlength{\minitwocolumn}
\begin{document}


\begin{flushright}
RIMS-1141 \\
\end{flushright}

\pagestyle{empty}

\begin{center}
{\large\bf The Exact Solution of Born-Infeld Theory \\
in Two Dimension\\
\vskip 1mm
}

\vspace{15mm}

Noriaki IKEDA
\footnote{ E-mail address:\ nori@kurims.kyoto-u.ac.jp } \\
Research Institute for Mathematical Sciences \\
Kyoto University, Kyoto 606-01, Japan
\end{center}
\date{}


\vspace{15mm}
\begin{abstract}
We obtain the exact operator solution of two-dimensional 
quantum Born-Infeld theory.
This theory has a Lagrangian density non-polynomial in the fundamental
fields.
So this analysis might shed some light on  
the analysis of 
non-perturbative effects of field theories.
We find the new exact soluble class of quantum field theories.
\end{abstract}

\newpage
\pagestyle{plain}
\pagenumbering{arabic}


\font\sc=cmr5 scaled\magstep1


\rm

\section{Introduction}

The Born-Infeld theory\cite{BI} is recently
investigated in the aspect of 
the string theory\cite{FraTse}$^{,}$\cite{Lei}.
However in this paper, we analyze this theory 
in view of quantum field theory for point particles.
It is necessary to analyze non-renormalizable field theories 
in detail, because non-renormalizability is one of the biggest
difficulties of quantum field theories. 
For example, four dimensional Einstein gravitational theory is not 
renormalizable in the conventional sense.

The non-polynomiality is a common
 feature of the gravitational theories.
which makes the analysis of the gravitational theory
difficult. 

The Lagrangian density of the Born-Infeld theory is non-polynomial in the
fundamental fields, and it is not renormalizable in
the sense of the naive power counting.
Thus the analysis of the Born-Infeld theory might give us a key to
investigate the above problems.

In this paper, we exactly solve two-dimensional Born-Infeld theory
in the light-cone gauge.
The method
to solve it is not the conventional perturbative method,
because in this theory, we cannot obtain the exact solutions by the usual
perturbation based on the free field.
The approach employed in our paper is based on the new method proposed 
in [4].

In [4], Abe and Nakanishi
have proposed the new method to solve quantum field theory by the
Heisenberg picture. 
The procedure is the following. 

First we calculate equal-time commutation relations of the fundamental
fields from the canonical commutation relations of canonical conjugate 
quantities.
From equal-time commutation relations and the equations of motion,
we set up the Cauchy problems for two-dimensional commutation relations.
By solving the Cauchy problems with operator coefficients,
we obtain the two-dimensional commutation relations of fundamental
fields.
Finally, we construct Wightman functions as they are
compatible with multiple commutation
relations for fundamental fields and energy positivity requirement.

We calculate all the exact multiple commutation relations and 
all n-point Wightman functions for the electromagnetic field.

\vskip50pt

\def\vev#1{{\langle#1\rangle}}
\def\calL{{\cal L}}
\def\brs{{\delta_B}}
\def\aplus{{A_{+}}}
\def\aminus{{A_{-}}}
\def\Dplus{{D^{(+)}}}
\def\Dplusin{{D^{(+)}_{<}}}
\def\wight#1{\langle#1\rangle}
\def\trun#1{\langle#1\rangle_{\hbox{\sc T}}}
\def\Da{{D}}
\def\Dplusa{{D^{(+)}}}
\def\Dplusina{{D^{(+)}_{<}}}

\section{The Solution}

The action of the Born-Infeld theory in two dimension is written as
\begin{eqnarray}
S & = & \int d^2 x {\cal L}, \nonumber \\ 
{\cal L} & = & \frac{1}{\lambda^2} \biggl[ - \sqrt{ - \det
(\eta_{\mu\nu} - \lambda F_{\mu\nu})} + \sqrt{ - \det \eta_{\mu\nu}}
\biggr] \nonumber \\
& = & \frac{1}{\lambda^2} \left[ - \sqrt{ 1 + \frac{\lambda^2}{2}
F_{\mu\nu} F^{\mu\nu}} + 1 \right],
\label{Act}
\end{eqnarray}
We rewrite the coupling constant as 
$\lambda^2 = 2 \kappa$ for simplicity.
Then we can rewrite (\ref{Act}) as
\begin{eqnarray}
{\cal L} = \frac{1}{2 \kappa} \left[ - \sqrt{1+ \kappa F_{\mu\nu}
F^{\mu\nu}} + 1 \right].
\label{Lag}
\end{eqnarray}
Expanding the Lagrangian density in power of $\kappa$, 
we obtain the Maxwell theory
at zeroth order:
\begin{eqnarray}
{\cal L} = - \frac{1}{4} F_{\mu\nu} F^{\mu\nu} + \cdots.
\end{eqnarray}
However the Lagrangian density becomes the infinite series of the
fundamental fields.
The mass dimension of $\kappa$ is $[M]^{-1}$.
Hence this theory is non-renormalizable 
in the naive sense.

In order to carry out canonical quantization, we fix the 
gauge of $U(1)$ symmetry. We take the light-cone gauge:
$\aminus = 0$, where 
\begin{eqnarray}
A_{\pm} = A_0 \pm A_1.
\end{eqnarray}
Then (\ref{Lag}) is written as 
\begin{eqnarray}
{\cal L} & = & \frac{1}{2 \kappa} \left[ - \sqrt{1 - 2 \kappa 
(\partial_{-} \aplus - \partial_{+} \aminus)^2}  + 1 \right]
+ B \aminus,
\label{Lagfix}
\end{eqnarray}
where $B$ is the Nakanishi-Lautrap field
and $x^\pm = x^0 \pm x^1$.
If we eliminate $A_{-}$ explicitly, (\ref{Lagfix}) becomes
\begin{eqnarray}
{\cal L} & = & \frac{1}{2 \kappa} \left[ - \sqrt{1 - 2 \kappa
(\partial_{-} A_{+} )^2}  + 1 \right].
\end{eqnarray}
The equations of motion are derived from (\ref{Lagfix}) as follows:
\begin{eqnarray}
&& \partial_{-} \left[ 
\frac{\partial_{-} \aplus}
{\sqrt{1 - 2 \kappa ( \partial_{-} \aplus )^2}}
\right] = 0, 
\label{eqA}
\\
&& \partial_{+} \left[ 
\frac{\partial_{-} \aplus}
{\sqrt{1 - 2 \kappa ( \partial_{-} \aplus )^2}}
\right] = B, \\
&& \aminus = 0.
\end{eqnarray}
The canonical conjugate momentum of $A_{+}$ is
\begin{eqnarray}
\pi_{\aplus} \equiv 
\frac{\partial {\cal L}}{\partial (\partial_0 A_+)}
= \frac{\partial_{-} \aplus}
{2 \sqrt{1 - 2 \kappa ( \partial_{-} \aplus )^2}}.
\label{pieq}
\end{eqnarray}

We calculate a solution which is
\begin{eqnarray}
{\sqrt{1 - 2 \kappa ( \partial_{-} \aplus )^2}} \neq 0.
\end{eqnarray}
We solve the equations of motion. 
From (\ref{eqA}), we can write
\begin{eqnarray}
\frac{\partial_{-} \aplus}
{\sqrt{1 - 2 \kappa ( \partial_{-} \aplus )^2}} = f(x^+),
\label{1int}
\end{eqnarray}
where $f(x^+)$ is a function depending only on $x^+$.
Then (\ref{1int}) is rewritten as
\begin{eqnarray}
\partial_{-} \aplus = \frac{f(x^+)}
{\sqrt{1 + 2 \kappa f^2(x^+)}}.
\label{Arepf}
\end{eqnarray}
So we can solve $\aplus(x)$ as
\begin{eqnarray}
\aplus(x) = \frac{f(x^+)} {\sqrt{1 + 2 \kappa f(x^+)^2}} x^- + g(x^+),
\label{aplusrep}
\end{eqnarray}
where $g(x^+)$ is a function depending only on $x^+$.
%
%
From (\ref{Arepf}), 
we obtain
\begin{eqnarray}
{\sqrt{1 - 2 \kappa ( \partial_{-} \aplus )^2}} 
= \frac{1}{\sqrt{1 + 2 \kappa f^2(x^+)}}.
\end{eqnarray}
$\kappa = \frac{\lambda^2}{2} > 0$.
Then if $f(x^+)$ is hermitian,
${\sqrt{1 - 2 \kappa ( \partial_{-} \aplus )^2}}$ in the action is not
zero. 

We can express 
$\pi_{\aplus}$ in terms of  $f(x^+)$ from (\ref{pieq}) and (\ref{1int}):
\begin{eqnarray}
\pi_{\aplus} = \frac{1}{2} f(x^+).
\label{pif}
\end{eqnarray}

In order to quantize the theory, we set up the canonical commutation
relations of the canonical quantities as follows:
\begin{eqnarray}
&& [ \pi_{\aplus}, \aplus ]|_{0} = - i \delta( x^1 - y^1), \nonumber \\
&& [ \aplus, \aplus ]|_{0}  = 0, \nonumber \\
&& [ \pi_{\aplus}, \pi_{\aplus} ]|_{0} = 0,
\label{cancom}
\end{eqnarray}
where $[,]|_{0} $ denotes the equal-time commutation relations at 
$x^0 = y^0$. 
If we substitute (\ref{aplusrep}) and (\ref{pif}) to (\ref{cancom}),
we obtain the equal-time commutation relations of $f$ and $g$.
Since $f$ and $g$ depend on only $x^+$,
two dimensional commutation relations 
of $f$ and $g$ are calculated as follows:
\begin{eqnarray}
&& [ f(x^+), f(y^+)] = [ g(x^+), g(y^+)] =0, \nonumber \\
&& [ f(x^+), g(y^+)] = - 2 i \delta( x^+ - y^+).
\label{fandg}
\end{eqnarray}
Since $\partial_- f = \partial_- g = 0$, 
$f$ and $g$ are the currents which generate the residual gauge
symmetries.
From (\ref{fandg}) and (\ref{aplusrep}), we can derive the
two-dimensional commutation relation of $\aplus$ as follows:
\begin{eqnarray}
[ \aplus(x), \aplus(y) ] &=& - 2 i
\frac{1}{(1+2 \kappa f(x)^2)^{\frac{3}{2}}}
(x^- - y^-) \delta( x^+ - y^+) \nonumber \\
&=& - 2 i [1 - 2 \kappa (\partial_{-} \aplus(x))^2 ]^{\frac{3}{2}}
(x^- - y^-) \delta( x^+ - y^+) \nonumber \\
&=& - \frac{i}{\pi}[1 - 2 \kappa (\partial_{-} \aplus(x))^2 ]^{\frac{3}{2}}
\Da(x - y),
\label{acomm}
\end{eqnarray}
where $\Da(x)$ is 
defined as 
\begin{eqnarray}
\Da(x) = 2 \pi x^{-} \delta(x^+).
\end{eqnarray}
%
We obtain an  infinite series when we expand (\ref{acomm}) in power of
the coupling constant 
$\kappa$.
We cannot determine the exact $\kappa$ dependence by the usual 
perturbation theory. 
We derive multiple commutation relations of $\aplus$
recursively:
\begin{eqnarray}
&& [ \cdots, [ \aplus(x_1), \aplus(x_2) ], \cdots, \aplus(x_n)] 
\nonumber \\
&& \quad = (- 2 i)^{n-1} 
\left[ \left( \frac{d}{d z} \right)^{n-1} \left( \frac{z}{\sqrt{1 
+ 2 \kappa z^2}} \right) \right] \Biggr|_{z = f(x_1)}
 (x_1^- - x_2^-) \nonumber \\
&& \qquad \times \delta( x_1^+ - x_2^+) \delta( x_1^+ - x_3^+) \cdots
\delta( x_1^+ - x_n^+), \nonumber \\
&& \quad = \left(- \frac{i}{\pi}\right)^{n-1} 
\left[ \left( \frac{d}{d z} \right)^{n-1} \left( \frac{z}{\sqrt{1 
+ 2 \kappa z^2}} \right) \right] \Biggr|_{z = f(x_1)}
\nonumber \\
&& \qquad \times \Da( x_1 - x_2) \partial_{-}^{x_1} \Da( x_1 - x_3) \cdots
\partial_{-}^{x_1} \Da( x_1 - x_n),
\label{multicomm}
\end{eqnarray}

Next, we derive the Wightman functions of this theory.
We set the vacuum expectation values of $f$ and $g$ as
\begin{eqnarray}
\wight{f(x^+)} &=& F(x^+), \nonumber \\
\wight{g(x^+)} &=& G(x^+),
\end{eqnarray}
where $F$ and $G$ are some c-number functions.
If these one-point functions are non-vanishing, 
Lorentz invariance is broken, but we dare to include nonzero
expectation values 
to consider general situations.
%
As is considered later,
if $F$ is a constant, the vacuum satisfies the subsidiary condition
which define the physical space.
Since
\begin{eqnarray}
[ f(x), f(y) ] = 0,
\label{fcomf}
\end{eqnarray}
we can trivially calculate $n$-point Wightman functions of $f$.
For example, the two-point Wightman function of $f(x)$ is calculated
as 
\begin{eqnarray}
\wight{f(x)f(y)} = \wight{f(x)}\wight{f(y)} = F(x) F(y).
\end{eqnarray}
Using the $n$-point functions of $f(x)$, we
obtain the one-point function of $\aplus$ as follows:
\begin{eqnarray}
\wight{\partial_{-} \aplus(x)} &=& 
\frac{F(x)}{\sqrt{1 + 2 \kappa F(x)^2}}, \nonumber \\
\wight{\aplus(x)} &=& \frac{F(x)}{\sqrt{1 + 2 \kappa F(x)^2}} x^{-}+ G(x),
\end{eqnarray}
Since two $\partial_{-} \aplus$'s commute mutually as is seen
from (\ref{Arepf}) and (\ref{fcomf}):
\begin{eqnarray}
[ \partial_{-} \aplus(x), \partial_{-} \aplus(y)] = 0, 
\label{dada}
\end{eqnarray}
we obtain the two-point truncated Wightman function of 
$\partial_{-} \aplus$
as
\begin{eqnarray}
\trun{\partial_{-} \aplus(x_1) \partial_{-} \aplus(x_2)} = 0.
\end{eqnarray}
Hence (\ref{acomm}) and the energy positivity\cite{AbeNak} requirement 
lead the
two-point Wightman function of $\aplus$ to 
\begin{eqnarray}
\trun{\aplus(x_1) \aplus(x_2)} & = & 
- \frac{1}{2 \pi} \sum_{i=1}^2
[ 1 - 2 \kappa \wight{(\partial_- \aplus(x_i))}^2 
]^{\frac{3}{2}} 
\Dplusa(x_1 - x_2) \nonumber \\
& = & 
- \frac{1}{2 \pi} \sum_{i=1}^2
\frac{1}{(1 + 2 \kappa F(x_i)^2 )^{\frac{3}{2}}}
\Dplusa(x_1 - x_2),  
\end{eqnarray}
where
\begin{eqnarray}
\Dplusa(x) = \frac{x^-}{x^+- i 0}.
\end{eqnarray}
From (\ref{multicomm}), we can calculate the $n$-point Wightman
functions of 
$\aplus$,
\begin{eqnarray}
&&\trun{\aplus(x_1) \aplus(x_2) \cdots \aplus(x_n)}
= \frac{1}{n(n-2)!} 
\sum^{n!}_{P(i_1, \cdots, i_n)}\left[ 
\left( \frac{d}{d Z} \right)^{n-1} \left( \frac{Z}{\sqrt{1 
+ 2 \kappa Z^2}} \right) \right]\Biggr|_{Z = F(x_{i_1})}
\nonumber \\
&& 
\qquad \qquad 
\times 
\left(- \frac{1}{\pi} \right)^{n-1} 
\Dplusina( x_{i_1} - x_{i_2}) 
\partial_{-}^{x_{i_1}} \Dplusina( x_{i_1} - x_{i_3}) \cdots
\partial_{-}^{x_{i_1}} \Dplusina( x_{i_1} - x_{i_n}),
\label{nwight}
\end{eqnarray}
%
%
where
\begin{eqnarray}
\Dplusina(x_i - x_j) = \left\{\begin{array}{ll}
\Dplusa(x_i- x_j), & \mbox{if $i<j$} \\
\Dplusa(x_j- x_i), & \mbox{if $i>j$} 
\end{array}
\right.
\end{eqnarray}
and $P(i_1, \cdots, i_n)$ is a permutation of $(1, \cdots, n)$.


The exact Wightman functions break the equations of motion in some
theories\cite{AN2}.
Therefore we discuss the consistency with the above solution
(\ref{nwight}) and the equations of motion.

We find that if a truncated Wightman function includes 
$\partial_{-}{}^{x_k} \aplus(x_k)$, 
it does not depend on $x_k{}^-$ from (\ref{nwight}).
And since $\partial_{-} \aplus$ commute mutually in two dimension and
the truncated $n$-point functions of $\partial_{-} \aplus$ are zero,
the Wightman functions which include any functional of
$\partial_{-} \aplus$ are non-singular at the same spacetime point.
Thus
\begin{eqnarray}
\left\langle \frac{\partial_-{}^{x_1} A_+(x_1)}
{\sqrt{1-2\kappa (\partial_-{}^{x_1} A_+(x_1))^2}}
A_+(x_2)\cdots A_+(x_n)\right\rangle, 
\end{eqnarray}
does not depend on $x_1{}^-$.
Therefore 
\begin{eqnarray}
\left\langle \partial_-{}^{x_1} \left[\frac{\partial_-{}^{x_1} A_+(x_1)}
{\sqrt{1-2\kappa (\partial_-{}^{x_1} A_+(x_1))^2}} \right]
A_+(x_2)\cdots A_+(x_n)\right\rangle =0, 
\end{eqnarray}
and we can confirm the Wightman functions is consistent with the
equation of motion.

We can set up the subsidiary condition to select the physical Fock
space as follows:
\begin{eqnarray}
B^{(+)}(x) |{\rm phys} \rangle = 0,
\label{subsidi}
\end{eqnarray}
where ${}^{(+)}$ is the positive frequency part of the field
$B$.
From the equations of motion, 
we obtain the following relation:
\begin{eqnarray}
B(x) = \partial_{+} f(x^+), 
\end{eqnarray}
so (\ref{subsidi}) is equivalent to the following one,
\begin{eqnarray}
\partial_+ f^{(+)}(x^+) |{\rm phys} \rangle = 0.
\end{eqnarray}

\vskip50pt

\section{Conclusion and Discussion}

We have exactly solved the quantum Born-Infeld theory 
in two dimension in the light-cone gauge. 
The usual perturbation method based on the free field does not work.
We calculated the exact multiple commutation relations and 
$n$-point Wightman functions by the non-perturbative method.
This method for solving  quantum theory will be useful to treat
non-perturbative effects of quantum field theories.

The generalization to the non-abelian Born-Infeld theory\cite{Hag} 
is straightforward.


The Schwinger model\cite{Sch}, which is exactly soluble, is an
important two-dimensional field theory.
So it may be interesting to analyze the Born-Infeld theory coupled with
fermion matters.

We can generalize the Lagrangian density (\ref{Lag}) to 
\begin{eqnarray}
{\cal L} = \Phi(F_{\mu\nu} F^{\mu\nu}),
\end{eqnarray}
where $\Phi(x)$ is a function.
If $\Phi(x)$ is satisfied with a certain condition,
the theory is also soluble
\footnote{The author thank to Prof.Nakanishi for pointing out it.}.
We will analyze the above generalized theory in detail at the next
paper. 

\begin{flushleft}
{\bf Acknowledgment}
\end{flushleft}
The author thank Prof.M.Abe and Prof.N.Nakanishi 
for discussions and comments about the present work. 
He express gratitude to Prof.N.Nakanishi
for reading the manuscript carefully.


\end{document}